\documentclass[rnote]{aa} 
\usepackage{epsfig}
\usepackage{rotating}
\usepackage{natbib}
\usepackage{graphicx}
\newcommand{\ein}{{\em Einstein}}

\newcommand{\ros}{{\em ROSAT}}

\newcommand{\chan}{{\em Chandra}}
\newcommand{\xmm}{{\em XMM-Newton}}
\newcommand{\vltn}{{\em Very Large Telescope}}
\newcommand{\vlt}{{\em VLT}}
\newcommand{\fors}{{\em FORS1}}
\newcommand{\forsn}{{\em FOcal Reducer/low dispersion Spectrograph}}
\newcommand{\gsc}{{\em GSC-2}}

\newcommand{\iraf}{{\em IRAF}}

\newcommand{\rxj}{RX\,  J0822.0$-$4300}

\begin{document}

\title{Deep optical observations of the central X-ray source in the Puppis A supernova remnant\thanks{Based on observations collected at ESO, Paranal, under Programme 78.D-0706(A)}}

\author{R. P. Mignani\inst{1}
\and
A. De Luca\inst{2}
\and 
S. Mereghetti\inst{2}
\and
P.A. Caraveo\inst{2}
}

\institute{Mullard Space Science Laboratory, University College London, Holmbury St. Mary, Dorking, Surrey, RH5 6NT, UK
\and
INAF, Istituto di Astrofisica Spaziale, Via Bassini 15, Milan, 20133, Italy
 }

\titlerunning{Optical observations of the Puppis A CCO}

\authorrunning{R.P. Mignani}
\offprints{rm2@mssl.ucl.ac.uk}

\date{Received ...; accepted ...}
\abstract{X-ray observations reveiled a group of radio-silent isolated
neutron stars (INSs) at the centre of young supernova remnants (SNRs),
dubbed central compact objects or CCOs, with properties different from
those of classical  rotation-powered pulsars. In at least  three
cases, evidence  points towards  CCOs being low-magnetized  INSs, born
with slow rotation periods, and  possibly accreting from a debris disc
of  material formed  out  of the  supernova  event. Understanding  the
origin  of the  diversity of  the CCOs  can shed  light  on supernova
explosion  and neutron star  formation models.   Optical/infrared (IR)
observations are crucial to  test different CCO interpretations. }{The aim
of our work is to perform a deep optical investigation of the CCO RX\, J0822.0$-$4300 in the
Puppis  A SNR, one of the most poorly understood in the CCO family. }{By
using  as  a  reference  the   \chan\  X-ray  coordinates of \rxj\,  we
performed deep optical  observations in the B, V and  I bands with the
\vltn\  (\vlt). }{We found no candidate optical counterpart within $3 \sigma$
of the computed \chan\ X-ray position down to $5 \sigma$ limits of B$\sim 27.2$, V $\sim
26.9$, and I $\sim 25.6$, the  deepest obtained in the optical band  for this
source.}{These limits confirm the non-detection of a companion brighter than an M5 dwarf. At the same time, they do not constrain optical emission from the neutron star  surface, while emission from the magnetosphere would require a spectral break in the optical/IR.  }

 \keywords{Stars: neutron, Stars: individual:  RX\, J0822.0$-$430}

   \maketitle

\section{Introduction}

Observations performed with the  \ein\ satellite lead to the discovery
of puzzling X-ray sources in young ($\sim$
2-10 kyears) supernova remnants (SNRs), i.e.  1E\, 1613$-$5055 in RCW
103  (Tuohy \&  Garmire  1980)  and 1E\,  1207$-$5209  in PKS\,  1209$-$52
(Helfand  \&  Becker  1984).    Their  location,  very  close  to  the
geometrical  centre of  the remnant, suggested  that they
were young  neutron stars formed by the  supernova explosions, despite
no pulsations being detected at radio and X-ray  wavelengths.  
About ten  of  these X-ray  sources,  dubbed central  compact
objects or CCOs (Pavlov et al.  2002), have been discovered to date.
The  CCO X-ray properties  markedly  differ  from  those of  young
rotation-powered INSs (see  De Luca 2008 for a  recent review). Their X-ray  spectra are not  purely magnetospheric, as  is the case, e.g. for the Crab pulsar,  but have
strong  thermal  components.  Furthermore, CCOs show no evidence of pulsar  wind nebulae
(PWN).  X-ray  pulsations have  been firmly detected  for two  CCOs, 1E\, 1207$-$5209  (P=424  ms;  Zavlin  et  al.   2000)  and  CXOU\,
J185238.6+004020    in    Kes    79    (P=105    ms;    Gotthelf    et
al. 2005).
Interestingly, in both  cases the measured upper limits  on the period
derivatives yielded spin  down ages which are a  factor of $\sim 10^3$
larger than  the SNR  age, as well  as dipole magnetic  fields $B <  5 \times
10^{11}$  G and  rotational energy  losses $\dot  E <  10^{32-33}$ erg
cm$^{-2}$ s$^{-1}$  (Gotthelf \& Halpern 2007; Halpern  et al.  2007).
One possibility is  that these CCOs were born  spinning close to their
present period and with very  low magnetic fields, enabling polar
cap  accretion  from  a  debris  disc  formed  out  of  the  supernova
explosion.   
 In the  optical/infrared (IR),
deep observations have  been performed only for a  handful of objects,
mostly resulting  in non-detections, which do not rule out the presence of a very low  mass companion star (M type or later),
or of a debris disc
(see De Luca et al.
2008,  and  references therein).

One of  the most interesting  CCOs is that  in the Puppis A  SNR (3700
years).  The  X-ray source (RX\,  J0822.0$-$4300), only hinted  at in \ein\
data (Petre et al.  1982), was then clearly detected by \ros\ (Petre et al.
1996). The  source X-ray  spectrum is predominantly
thermal  (Hui \&  Becker  2006) and well fitted  by a two-component
blackbody model. 
The  corresponding hydrogen column density ($N_H \sim
4.54  \times  10^{21}$  cm$^{-2}$)  is  consistent with  the  2.2  kpc
distance  derived  from  neutral  hydrogen  observations  of  the  SNR
(Reynoso  et  al.   2003),  which yield a 0.5--10 keV X-ray luminosity of $L_{X}\sim 2-5 \times 10^{33}$ erg s$^{-1}$ for \rxj. Recently, a period at $\sim 112$ ms has been reported from \xmm\  observations by Gotthelf \& Halpern (2009).
The  first  optical  observations  of  RX\,
J0822.0$-$4300 yielded upper limits of B$\sim 25$ and R$\sim23.6$
on  the  source  brightness  (Petre  et al.   1996) while new  optical/IR
observations  (Wang et  al.  2007) did  not reveal  any candidate
counterpart down to B$\sim$ 26.5 and K$_S$ $\sim$ 20.1

Here we present  deeper optical observations  of
 RX\, J0822.0$-$4300,  performed with the \vltn\ (\vlt)  of the European
 Southern  Observatory  (ESO).  Observations  and  data reduction  are
 described in Sect.  2, while  the results are presented and discussed
 in Sect. 3 and 4, respectively.

\section{Observations}

\subsection{Observations and data reduction}

Optical imaging  observations of  RX\, J0822.0$-$4300 were  performed in
service  mode with  the  \vlt\  Kueyen telescope  at  the ESO  Paranal
observatory in November  2006 and   March 2007
(see Table 1).  The latter observations were performed right after
the  recoating of  the  telescope  main mirror on March 2nd 2007.  We  used the  \forsn\
(\fors),  a multi-mode camera  for imaging  and long-slit/multi-object
spectroscopy. At  the time  of our observations  the camera  was still
equipped  with the  original single  chip, four  port 2048$\times$2084
Tektronik  CCD detector.   We selected  the standard  resolution mode,
with a 0\farcs2  pixel size, and a field of  view of 6$\arcmin8 \times
6\arcmin8$.  Sequences of  560 s exposures were obtained through
the Bessel B,  V and I filters, for total  integration times of 14000,
2800,   and  11200  s,   respectively.   In   order  to   avoid  light
contamination at  the target  position, bright stars  in the  field of
view were  masked using the \fors\ occulting  bars.  

\begin{table}[h]
\begin{center}
  \caption{Log of  the \vlt/\fors\  BVI observations of the Puppis A CCO (RX\, J0822.0$-$4300). Columns report the observing date, the filter name, the total integration time, the average seeing and airmass in each data set. }
\begin{tabular}{lcccc} \\ \hline
yyyy-mm-dd & Filter & T (s) & Seeing    & Airmass	\\ \hline 
2006-11-26 &  I & 2800 & 0.88 & 1.09  \\
2007-03-18 &  B & 2800 & 0.98 & 1.06  \\
2007-03-20 &  B & 5600 & 0.84 & 1.07  \\
           &  V & 2800 & 0.71 & 1.14  \\
2007-03-23 &  I & 8400 & 0.71 & 1.22  \\
2007-03-24 &  B & 2800 & 1.22 & 1.05  \\  
2007-03-25 &  B & 2800 & 0.89 & 1.05  \\        
\hline 
\end{tabular}
\label{data}
\end{center}
\end{table}

Observations were carried out with airmass better  than $\sim  1.3$, and  $\sim 0\farcs8$
seeing.   Sky  conditions  were photometric (or close  to) for all nights, with  the exception of that
of November 26th  2006.  For this reason, and due  to the limited amount
of  exposure time,  in the following we do not  make use  of this  observation.  
 Bias,  twilight  flat--fields,  and
observations of standard stars  from the Landolt fields (Landolt 1992)
were obtained as part of the \fors\ science calibration plan.  For the
V-band observations no standard star observations were acquired on the
night of March 21st.  For  photometric calibration purposes we thus used
the  closest-in-time available  standard star  observations,  taken on
March 10th.

Data were reduced (bias  subtraction, flat--field correction, pre- and
overscan   trimming)    using   the   ESO    \fors\   data   reduction
pipeline\footnote{http://www.eso.org/observing/dfo/quality/FORS1/pipeline}.
We applied the photometric  calibration using the  night zero
points    available   in   the    instrument   data    quality   control
database\footnote{http://www.eso.org/observing/dfo/quality/}.   Airmass
correction  was  applied using  the  extinction  coefficients for  the
Paranal Observatory.  For the March 21st  V-band data set the  zero point  trending
plots ensure that the used closest-in-time zero point of March 10th, i.e. measured after the telescope recoating, does
not introduce  systematic effects in our  photometric calibration.  We
converted the database zero points, computed in  units of electrons/s,
in  units of  ADU/s by  applying the  corresponding electrons--to--ADU
conversion  factors.  For  each dataset,  we finally  used  the \iraf\
tasks {\tt  drizzle} and {\tt  imcombine} to align and co--add all  exposures taken  with the  same
filter,  as well as to mask bad pixels and filter cosmic ray hits.

\begin{figure}
\centering 
\includegraphics[height=8.5cm,angle=0,clip]{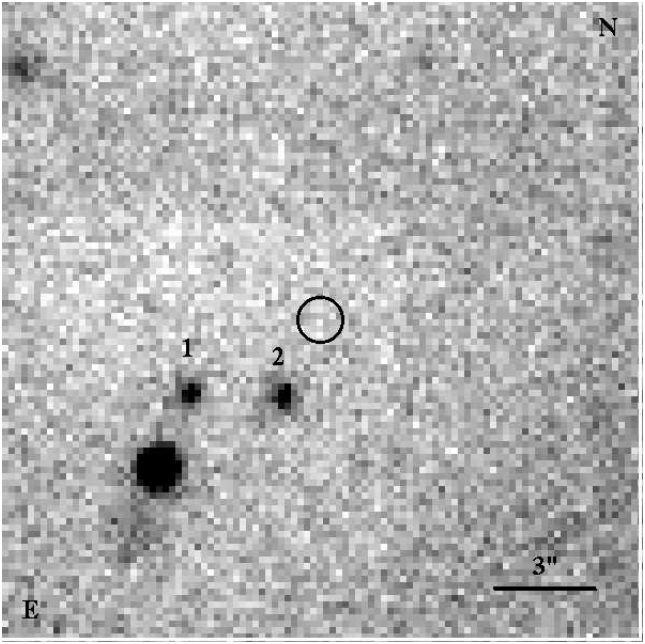} 
  \caption{$20\arcsec \times  20\arcsec$ cutout of  the co--added
\fors\ B-band image (14000 s)  of  the RX\, J0822.0$-$4300
field.   The CCO  position is  marked by the  circle 
(0\farcs7  radius, $3  \sigma$  confidence level).
The two objects  detected closest to the \chan\ position,
also  detected  in the  optical  images of  Wang  et  al. (2007),  are
labelled.  }
\label{fors1}       
\end{figure}

\subsection{Astrometry calibration}

The astrometric  calibration was computed on the  \fors\ B-band image,
taken  as  a reference,  using  the  position  and coordinates  of  50
well-suited stars (e.g.  not saturated,  or too faint, or too close to
the CCD edges,  and evenly distributed in the  field of view) selected
from  the  \gsc\  version  2.3  (Lasker  et  al.   2008).   The  pixel
coordinates  of  the  \gsc\  stars  were  measured  by  fitting  their
intensity profiles  with a Gaussian function using  the dedicated tool
of  the {\em  Graphical  Astronomy and  Image  Analysis} ({\em  GAIA})
interface\footnote{star-www.dur.ac.uk/$\sim$pdraper/gaia/gaia.html}.     The
transformation  between  the detector  and  sky  coordinates was  then
computed      using     the      {\em     Starlink}      code     {\tt
ASTROM}\footnote{http://star-www.rl.ac.uk/Software/software.htm} which
is based on  higher order polynomials and accounts  for unmodeled CCD
distorsions.  
The  rms  of  our  astrometric  solution  is  $\delta r \approx
0\farcs15$, accounting for the rms of the fit in the right ascension and declination components,
to which we added in quadrature the 
 uncertainty   in the 
registration of the \fors\ image on the \gsc\ reference frame.  Following Lattanzi  et al. (1997),  we estimated it as
$\sqrt 3 \times \sigma_{GSC} /  \sqrt N_{s}$, where the $\sqrt 3$ term
accounts for  the free parameters ($x$-scale,  $y$-scale, and rotation
angle)  in the  astrometric fit,  $\sigma_{GSC}=0\farcs3$ is  the mean
positional error of  the \gsc\  coordinates  (Lasker et  al. 2008)  and
$N_{s}$   is  the   number   of  stars   used   for  the   astrometric
calibration. The uncertainty on the reference star centroids is below
0\farcs01  and  was  neglected.   We  also  added  in  quadrature  the
0\farcs15  uncertainty (Lasker et al. 2008)  on the  link of  the \gsc\  to the
International  Celestial Reference  Frame (ICRF).   Thus,  the overall
accuracy of the \fors\ astrometry is 0\farcs22 ($1\sigma$).

\section{Results}

As a reference for the  RX\,  J0822.0$-$4300 position we used the \chan\ coordinates reported in Winkler  \&  Petre (2007), i.e.  $\alpha  (J2000)=08^h  21^m 57.355^s$,  $\delta  (J2000)=-43^\circ 00\arcmin 17\farcs17$ (epoch 2005.32; MJD= 53485), with a $1 \sigma$ uncertainty of $\pm 0\farcs01$ per each coordinate.  To register the
 RX\,  J0822.0$-$4300 position on  our \fors\  B-band image
(epoch 2007.23;  MJD=54184) we then  accounted for its \chan\ proper motion of 165 $\pm$ 25  mas yr$^{-1}$ (Winkler  \& Petre
2007), obtaining $\alpha  (J2000)=08^h  21^m 57.309^s$,  $\delta  (J2000)=
-43^\circ 00\arcmin 17\farcs28$.
We finally estimated the  overall uncertainty on the RX\, J0822.0$-$4300
position  at  the  epoch  of  our \fors\  observations  by  adding  in
quadrature the positional error on the \chan\ coordinates at the reference
epoch  ($0\farcs015$), the  coordinate uncertainty  due to  the proper
motion  extrapolation  ($\sim  0\farcs048$),  and  the  error  of  our
astrometric calibration ($\sim 0\farcs22$).  This yields to an overall
uncertainty of  0\farcs23 (1$\sigma$).  We  note that both  the \chan\
coordinates and  the \gsc\ ones, used for  the astrometric calibration
of our \fors\ images, are all linked  to the ICRF so that we exclude any
systematic offset in our astrometry.

Figure  1 shows the  computed \chan\ position of RX\, J0822.0$-$4300
overlaid  on   the  co-added  \fors\  B-band   image.   We note that our error circle is offset by $\sim 1\farcs3$ from that shown in Figure 1 of Wang et al. (2007). This is due to the X-ray source proper motion between the epoch of our \vlt\ observation (March 24th 2007) and the \chan\ one  (January 1st 2000) from which they derived the reference coordinates for  their {\em Magellan} observation (April 5th 2003).
No  candidate
counterpart is  detected within a  conservative $\sim 3  \sigma$ error
circle.  
We  have computed  the  flux  of  the field  objects  through
customized aperture photometry.  The closest
objects, marked as 1 (B=25.6) and 2 (B=25.8) in Figure 1, are detected at
$\sim  2\farcs6$  and  $4\farcs7$   southeast  of  the  CCO  position,
respectively.   Their measured  offests are  well beyond  any possible
residual  uncertainty   in  our   astrometry,  which  rules   out  any
association with the CCO.  No candidate counterpart is detected either
in the  V-band image   or in the  I-band one.  We thus conclude that  the optical counterpart of the Puppis
A CCO  is unidentified. 
 We  estimated the detection  limits of our images  by  extrapolating  the  fluxes of  the  faintest  objects
detected close  to the  RX\, J0822.0$-$4300 position.   To this  aim, we
run the object  detection  using the  {\em SExtractor}  program
(Bertin  \& Arnouts  1996) and  we computed  their  magnitudes through
customized  aperture photometry.   We  thus derived  $5 \sigma$  upper
limits on the optical  brightness of the RX\, J0822.0$-$4300 counterpart
of B$\sim 27.2$, V $\sim 26.9$, and I $\sim 25.6$.

\section{Discussion and conclusions}

\begin{figure}
\centering 
\includegraphics[height=8.5cm,angle=-90,clip]{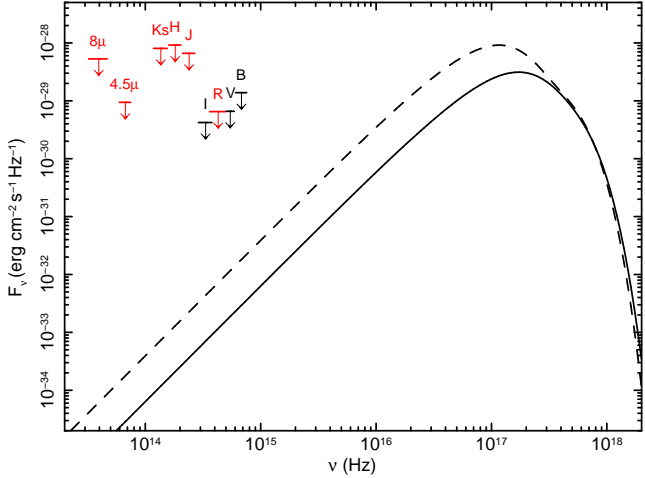} 
  \caption{Extrapolation in the optical/IR domain of the X-ray spectrum of  RX\,
J0822.0$-$4300.  Both curves correspond to a double blackbody best-fit spectrum  (Hui \& Becker 2006a) with (i) $T_{1} = 2.61 \times 10^{6}$ K,$T_{2} = 5.04 \times 10^{6}$ K,  and emitting radii $R_{1} = 3.29$ km, $R_{2}=0.75$ km (solid line)  and   (ii) $T_{1} = 1.87 \times 10^{6}$ K, $T_{2} = 4.58\times10^{6}$ K, $R_{1} = 10$ km, $R_{2}=1.09$ km (dashed line). The extinction-corrected optical/IR flux upper limits (Wang et al. 2007 and  present work) are plotted in black and red, respectively. }
\label{spec }       
\end{figure}

Our  BVI flux limits  confirm the non-detection of a companion star earlier than an M5 dwarf,  as suggested by the previous optical and near/mid-IR flux upper limits of Wang et al. (2007).   Similar conclusions were recently reported for other CCOs (see De Luca 2008 and references therein).
The available upper limits can also be used to constrain the optical/IR spectrum from the neutron star or from a possible fallback disc.  In Figure 2 we plotted the extinction-corrected near/mid-IR and optical spectral flux upper limits together with the extrapolations of the best fits to the X-ray spectrum computed using 
double blackbody models 
with the hydrogen column density $N_{H}$ left as a free parameter  (Hui \& Becker 2006). 
We computed  the  interstellar extinction correction  in the  JHK$_s$ and BVRI  bands  from  the
$N_{H}$ derived  from the X-ray spectral fits applying the relation of Predehl \& Schmitt (1995) and the extinction  coefficients of Fitzpatrick (1999).  For the mid-IR bands the interstellar extinction was computed from the $A_V$ using the relations of Chiar \& Tielens (2006).  For simplicity, we plotted only the extinction-corrected spectral fluxes corresponding to an $N_H = 4.54 \times10^{21}$ cm$^{-2}$ derived from the first double blackbody spectral fit. 
As seen,  the X-ray spectra extrapolations are at least two orders of magnitude below the optical/IR upper limits, even in the more favorable case of thermal emission from the bulk of the neutron star surface (dashed line in Figure 2).  On the other hand,  for a blackbody plus power--law model ($T_{1}= 8.57 \times10^{6}$ K; $\Gamma=4.29$) the X-ray spectrum extrapolation falls about four orders or magnitudes above the optical/IR upper limits.  Thus, we can not constrain optical/IR emission from the neutron star surface, while emission from the magnetosphere would obviously require a break in the source optical--to--X-ray power-law spectrum.  We note that the spectral break implied by the broken power-law model fit to the X-ray spectrum ($\Gamma_{1}= 3.61$; $\Gamma_{2}=5.29$) would also over-predict the optical emission of \rxj\ by an amount similar to the blackbody plus power-law model.  However, detection  of optical magnetospheric emission would require an energy output and an optical emission efficiency comparable to those of the Crab pulsar.

The presence of a fallback disc can not be ruled out by our data.  For the assumed 0.5--10 keV X-ray luminosity of  $L_{X}\sim 2-5 \times 10^{33}$ erg s$^{-1}$, the disc emission would be dominated by the reprocessed X-ray flux (Perna et al. 2000). The X-ray luminosity of \rxj\  is within a factor of 2 from  that of the CCO  1WGA\,   J1713$-$3949 in G347.3-0.5, i.e.  $L_{X}\sim 2.8 \times 10^{33}$ erg s$^{1}$ at a distance of 1.3 kpc (see Mignani et al. 2008 and references therein). This would imply, for the same set of disc model parameters, i.e. the disc inner and outer radii, the disc accretion rate, and the viewing geometry  (Perna et al. 2000),  a comparable disc IR luminosity. 
Thus, the optical/IR spectral flux upper limits of \rxj\ are still consistent with the presence of, e.g. an undetected disc with outer radius smaller than  $\approx 1 R_{\odot}$.  

The present upper limits on the \rxj\  optical emission can only be marginally improved with the \vlt, or with other  8m-class telescopes.  On the other hand,  much deeper IR observations of \rxj\ are still feasible with the \vlt\  and could  help to  further constrain the presence of a fallback disc.

\begin{acknowledgements}
RPM acknowledges STFC for support through its Rolling Grant programme. The authors thank  Emanuela Pompei for support with the \vlt\ observations and Rosalba Perna for useful comments.
\end{acknowledgements}

\end{document}